\documentclass[aps,prl,twocolumn,superscriptaddress,groupedaddress,secnumarabic,floatfix, nofootinbib,tightenlines,nobibnotes, aps, prl, 11pt,showkeys]{revtex4-2}  
\usepackage{blindtext}
\usepackage{threeparttable}
\usepackage{tikz}
\usepackage{spverbatim}
 \usepackage[font=footnotesize,labelfont=bf]{caption}
 \usepackage{ragged2e}
 \justifying
 \usepackage{rotating}
 
\usepackage[section]{placeins}
 \usepackage{subcaption}
\usepackage{listings}
\usepackage{mathtools}

\usepackage{rotating}
\usepackage{graphicx}  
\usepackage{dcolumn}  
\usepackage{amsmath}
\usepackage{multirow}
\usepackage{bm}   
 \usepackage[colorlinks=true,linkcolor=blue]{hyperref}
\usepackage{amssymb}   
\usepackage{lipsum}
\usepackage{tablefootnote}
\usepackage{hhline}
\usepackage{booktabs}
\usepackage{tabularx}
\usepackage{arydshln}
\usepackage{xifthen}
\usepackage{float}
\usepackage{pifont}
\usepackage{fancyhdr}

\begin{document}
	\title{Effect of hydrostatic strain on the mechanical properties and topological phase transition of bi-alkali pnictogen NaLi$_{2}$Bi }
	\author{Seyed Mohammad bagher Malek Hosseini}
	\affiliation{Department of Physics, Vali-e-Asr University of Rafsanjan, Rafsanjan, Iran}
	\author{Shahram Yalameha}
\affiliation{Faculty of Physics,  University of Isfahan, 81746-73441, Isfahan, Iran.}

	\begin{abstract}
		The bi-alkali pnictogens have attracted significant attention for optoelectronic and photocathodic device applications. However, in most of the compounds belonging to this family, there has been less effort put into investigating the mechanical properties and topological phase transitions (TPT) of the compounds. Here, in the framework of density functional theory, the mechanical properties and topological phase transition of NaLi$_{2}$Bi under hydrostatic pressures are investigated. Elastic constants and phonon calculations have shown the mechanical and dynamical stability of this compound under hydrostatic tension and compression. The analysis of the elastic constants show that the NaLi$_{2}$Bi in the equilibrium state is an \textit{auxetic} material with a negative Poisson's ratio of -0.285, which changes to a material with a positive Poisson's ratio under hydrostatic tension. Meanwhile, Poisson's ratio and Pugh ratio indicate that this compound has brittle behavior and maintains it under hydrostatic pressures. The calculated results of the band structure within the generalized gradient approximation (GGA) (Tran-Blaha modified Becke-Johnson exchange potential approximation (TB-mBJ)) show that NaLi$_{2}$Bi is a nontrivial topological material (trivial topological material). It was found that hydrostatic compression (tension) in the GGA (TB-mBJ) approach leads to a transition from a nontrivial (trivial) to a trivial (nontrivial) topological phase for this compound. Moreover, the calculated Wannier charge centers confirm the TPT. Identifying the mechanisms controlling the \textit{auxetic} behavior and TPT of this compound offers a valuable feature for designing and developing high-performance nanoscale electromechanical and spintronic devices.

	\end{abstract}
 		
	\maketitle
	
\section{I. INTRODUCTION}

In general, bi-alkali pnictogens have the formula XY$_{2}$Z, in which X and Y represent alkali metals and Z represents group 7 elements. These compounds comprise an interesting class of semiconducting compounds that have technological importance in optoelectronics, sensing, and photocathodic devices, due to their favorable electronic properties \cite{ref1}. It is expected that these compounds will provide high-intensity X-ray electron Lasers in the future \cite{ref2,ref3,ref4}. In addition, these compounds are used in electron radiative devices because of their feasible work function \cite{ref5}. Recently, some of these compounds have attracted the attention of researchers due to their narrow band gap and strong spin-orbit coupling (SOC) \cite{ref6,ref8}.

From the experimental point of view, bi-alkali pnictogens have also been investigated. Using the in-situ X-ray diffraction method, Ruiz \textit{et al.}, \cite{ref9} investigated the crystalline structure of multi-alkali photocathodes. Also, Cultrera \textit{et al.}, \cite{ref10} have synthesized and characterized top-ordered bi and multi-alkali pnictogen photocathodes and involved the Rb element in their chemical structure. Feng \textit{et al.}, \cite{ref11} studied the effect of surface roughness on the emission of alkali penictogen cathodes in electron guns and developed K-Cs-Sb cathodes using co-deposition. Ding \textit{et al.} \cite{ref12} showed that co-evaporation is the best growth method for KCs$_{2}$Sb photocathodes because it produces films with a smooth surface, high quantum efficiency, and proper crystal structures. Schubert \textit{et al.}, \cite{ref13} investigated the bi-alkali pnictogen photocathodes for high-brightness accelerators.

From the theoretical perspective, the major attention on the bi-alkali pnictogens is to investigate their electronic and structural properties. However, there are also reports on the optical, elastic, and topological properties of a number of bi-alkali pnictogen compounds. Khan \textit{et al.}, \cite{ref14} reported the structural and optoelectronic properties of bi-alkali pnictogen compounds XY$_{2}$Z (X and Y = Li, Na, K, Rb, Cs; Z = N, P, As, Sb, Bi). Amador \textit{et al.}, \cite{ref15} described the electronic structure and optical properties of cubic Na$_{2}$KSb and hexagonal NaK$_{2}$Sb compounds. Murtaza \textit{et al.}, \cite{ref16} investigated the structural, elastic, electronic, and optical properties of some bi-alkali antimonides. Cocchi \textit{et al.}, \cite{ref17} studied the electronic and optical properties of the CsK$_{2}$Sb compound. Kalarasse \textit{et al.}, \cite{ref18} reported the electronic and elastic properties of the alkali pnictide compounds Li$_{3}$Sb, Li$_{3}$Bi, Li$_{2}$NaSb, and Li$_{2}$NaBi in the equilibrium state. Jin \textit{et al.}, \cite{ref19} investigated the electronic structure, doping effect, and topological signature in hexagonal compounds Li$_{3-x}$Na$_{x}$M (\textit{x} = 3, 2, 1, 0; M = N, P, As, Sb, Bi). Sklyadneva \textit{et al.}, \cite{ref8} reported that the KNa$_{2}$Bi compound is nontrivial topological material. Yalameha \textit{et al.}, \cite{ref7} investigated that the crystal structure of KNa$_{2}$Sb was also a topological insulator (TI) under hydrostatic pressure. Furthermore, they have recently shown that Rb(Na, K)$_{2}$Bi and Cs(Na, K)$_{2}$Bi compounds can take a diverse set of topological states by strain-engineering \cite{ref20,ref21}. They also investigated the mechanical and anisotropic elastic properties of Cs(Na, K)$_{2}$Bi compounds under hydrostatic pressure \cite{ref22}. Very recently, Song \textit{et al.} \cite{ref48} comprehensively investigated the thermal transport and thermoelectric properties of the bi-alkali bismuthide compound NaLi$_{2}$Bi.

So far, no comprehensive study has been devoted to examining the mechanical, anisotropic elastic, and topological properties of the cubic bi-alkali pnictogen compound NaLi$_{2}$Bi. In this work, the effect of hydrostatic tension and compression on mechanical properties (\textit{e.g.}, Young's modulus, Poisson's ratio, Pugh ratio, Cauchy pressure, etc.) and the TPT of this compound have been investigated. The mechanical and dynamic stability of the compound under the studied pressures has also been studied. It should be noted that in this composition, the TPT point in PBE-GGA (from nontrivial to trivial state) and TB-mBJ (from trivial to nontrivial state) approaches and their critical pressures have been observed and investigated. Even though this compound does not have any experimental reports, it is hoped that, based on these results, experimental research on this compound can be conducted.

\section{II. Computational details}
In this paper, the density functional theory (DFT)-based \textsc{Wien2k} package \cite{ref23} has been used to explore various physical properties of the titled compound. This package implements the full potential linearized augmented plane wave method (FP-LAPW) \cite{ref24}. In this work, the generalized gradient approximation (GGA) of the Perdew-Burke-Ernzerhof (PBE) exchange-correlation functional is used \cite{ref25}. The Tran-Blaha modified Becke-Johnson exchange potential approximation (TB-mBJ) \cite{ref49} calculations were also conducted in order to obtain improved band gaps. The important parameters G{\scriptsize max} (\textit{i.e.}, the potential and charge density Fourier expansion parameter) and R{\scriptsize MT}×K{\scriptsize max}, where R{\scriptsize MT} is the smallest muffin-tin (MT) sphere radii and K{\scriptsize max} is the largest reciprocal lattice vector used in the plane wave expansion, are set to 15.0 Bohr$ ^{-1} $ and 9.5 (Ry)$ ^{1/2} $, respectively. In the calculations of this compound, a 13×13×13 \textit{\textbf{k}}-mesh of points is used to achieve self-consistency. The phonon dispersion is calculated using the \textsc{Phonopy} package \cite{ref26}. In \textsc{Phonopy}, first-principles phonon calculations are performed using the finite displacement method \cite{ref27}. So, we used a supercell model and the finite displacement approach with a 2×2×2 conventional unit cell supercell and atomic displacement distance of 0.01 Å. \textsc{Irelast} \cite{ref28} and \textsc{ElATools} \cite{ref29} codes are used for elastic constants calculations and their analysis, respectively. \textsc{Wannier90} \cite{ref30} and \textsc{Z2pack} \cite{ref31} packages have been used to calculate Wannier charge centers (WCC). It is noteworthy that the hydrostatic pressures (tension- and compression-type) are evaluated using the volume ratio V/V{\scriptsize 0}, where V{\scriptsize 0}  is the equilibrium volume of the primitive unit cell. In all calculations presented in this study, spin-orbit coupling (with full relativistic effect) has been considered unless otherwise stated.
 
\section{III. Results and discussion}
 \subsection{III.A Structural properties, stability conditions}
 \begin{figure*}
 	\includegraphics[scale=1.0]{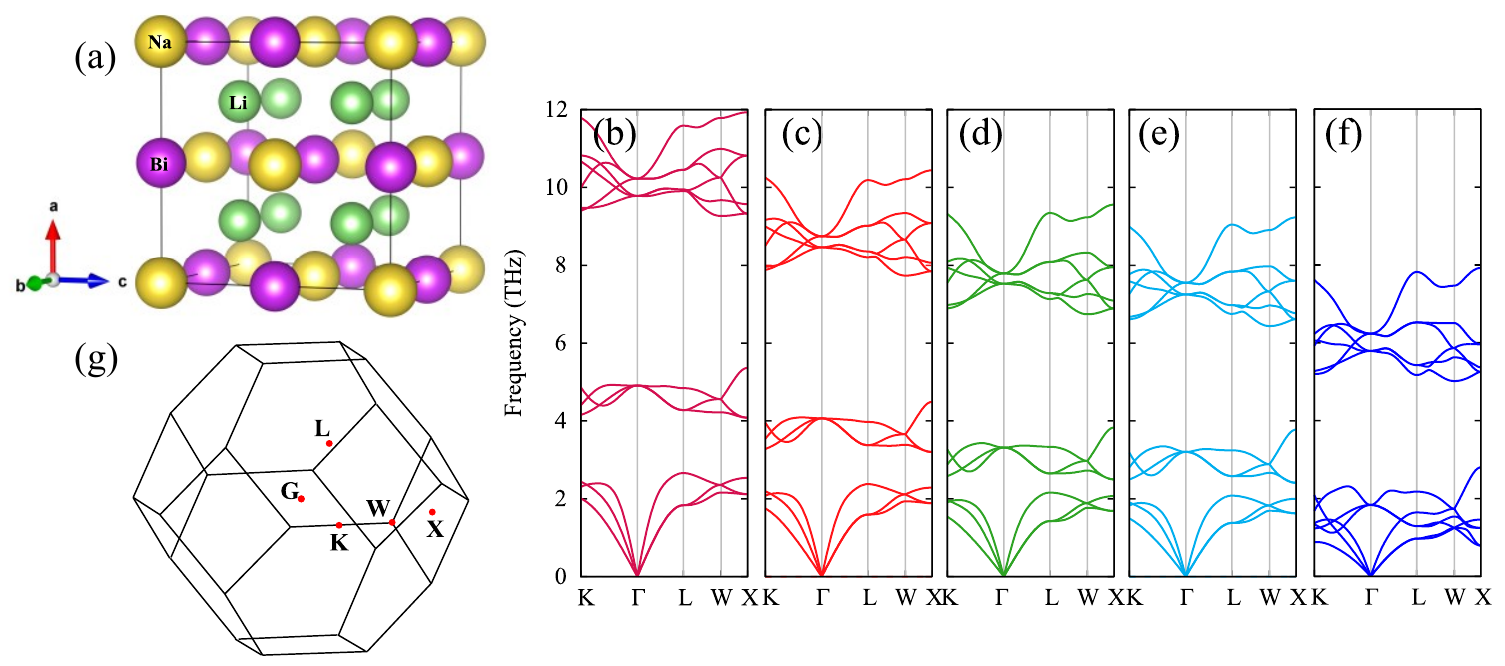} 
 	
 	\caption{\label{fig:wide_1}(a) Conventional crystal structure of bi-alkali pnictide NaLi$_{2}$Bi with space group Fm-3m (No. 225). The atomic positions in a unit cell are: Na atoms at (0, 0, 0), Bi atoms at (0.5, 0, 0) and Li(1)/Li(2) atoms at (0.75, 0.25, 0.25)/(0.25, 0.75, 0.75). Phonon dispersion curves along K-$\Gamma$-L-W-X high-symmetry points at (b) V/V$_{0}$ = 0. 930, (c) V/V$_{0}$ = 0.940, (d) V/V$_{0}$ = 1.0, (e) V/V$_{0}$ = 1.060 and (f) V/V$_{0}$ = 1.070. The phonon dispersion exhibits no imaginary frequency (negative frequency), confirming the dynamic stability of these compounds. (g) The first 3D Brillouin zone and the high-symmetry points. Note that here the meaning of the G (Gamma) point is the same as $\Gamma$ point.}
 	
 \end{figure*}
The crystal structure of the NaLi$_{2}$Bi compound is cubic with space group Fm-3m (No. 225), and the schematic conventional crystal structure of this compound is shown in Fig. ~\ref{fig:wide_1}(a). The atomic positions in a unit cell for Li(1)/Li(2), Na, and Bi atoms are (0.75, 0.25, 0.25)/(0.25, 0.75, 0.75), (0, 0, 0), and (0.5, 0, 0), respectively. A list of the results of first-principles calculations of the structural and elastic properties of this compound along with available theoretical lattice constants, is provided in Table~\ref{Tab_1}. By fitting the Murnaghan equation of state to the total energy as a function of volume, the equilibrium lattice parameter (\textit{a}$_{0}$) can be determined \cite{ref32} (see Fig. \textcolor{red}{1S} of Supplementary Information). The calculated equilibrium lattice constant (7.021 Å) is in good agreement with theoretical predictions (see Table ~\ref{Tab_1}). 
  
There are only three independent elastic constants for materials with cubic symmetry: C$_{11}$, C$_{12}$, and C$_{44}$. As can be seen in Table~\ref{Tab_1}, the calculated elastic constant values at the equilibrium state (V/V$_{0}$ = 1.0) in this work are consistent with other theoretical results. In order for a cubic crystal system to be mechanically stable, it needs to satisfy the following Born-Huang criteria \cite{ref33},
\begin{equation}
	C_{44} > 0, C_{11}-C_{12}>0, C_{11}+2C_{12}>0;
\end{equation}
According to Table~\ref{Tab_1} and Eq. 1, all elastic constants at the studied pressures of this compound satisfy the criteria of mechanical stability. In the equilibrium state and for hydrostatic tension (V/V$_{0}$ = 1.06 and 1.07) and compression (V/V$_{0}$ = 0.930 and 0.940) values, phonon dispersion curves for NaLi$_{2}$Bi along the high-symmetry points in the Brillouin zone (see Fig.~\ref{fig:wide_1}(g)) are plotted in Figs.~\ref{fig:wide_1}(b-f). It reveals that the NaLi$_{2}$Bi compound is dynamically stable at the studied pressures and equilibrium state since there are no modes with negative frequency values. To find the thermodynamical stability of the NaLi$_{2}$Bi compound, the Formation energies (\textit{E}$_{f}$) of volume ratios V/V$_{0}$ = 0.93, 0.94, 1.0, 1.06, and 1.07 are calculated. Table \textcolor{red}{1S} shows the calculated values of the \textit{E}$_{f}$ per formula unit. In this table, the calculated values of \textit{E}$_{f}$ per formula unit are negative, indicating this compound in the studied pressures is thermodynamically stable. The thermal stability of this compound has been recently investigated by Song \textit{et al.} \cite{ref48}. Thus, the NaLi$_{2}$Bi compound is stable and can be synthesized in the laboratory.

\begin{table*}
	\centering
	
	\caption{Calculated lattice constants (\textit{a}$_{0}$), elastic constants (C$_{ij}$), Cauchy pressure (C$_{P}$), Poisson's ratio ($\nu$), Pugh ratio (B/G), machinability index ($\mu_{M}$) and universal anisotropy index (A$^{U}$) relative to volume ratios (V/V$_{0}$) for NaLi$_{2}$Bi compound. It should be noted that the equivalent pressures are specified with each of the volume ratios.}
	\label{Tab_1}
	\begin{tabular}{cccccccccccc} 
		\hhline{============}
V/V$_{0}$                    & \begin{tabular}[c]{@{}c@{}}Pressure  (GPa)\end{tabular} & \begin{tabular}[c]{@{}c@{}}a$ _{0} $  (Å)\end{tabular}                 & \begin{tabular}[c]{@{}c@{}}C$_{11}$\\ (GPa)\end{tabular}        & \begin{tabular}[c]{@{}c@{}}C$_{12} $\\ (GPa)\end{tabular}      & \begin{tabular}[c]{@{}c@{}}C$ _{44} $\\ (GPa)\end{tabular}      & \begin{tabular}[c]{@{}c@{}}C$_{P}$\\ (GPa)\end{tabular} & $\nu$      & B/G    & $\mu_{M}$     & A$^{U}$     & A$^{E}$   \\

		\hhline{============}
		0.940         & 1.7                                       & 6.887                                                                                                                   & 50.48                                                                               & 19.89                                                                               & 42.35                                                                               & -22.46            & 0.1433              & 1.6527       & 1.2434               & 1.3562                        & 2.20                           \\ 
		\hhline{------------}
		0.955         & 1.0                                       & 6.922                                                                                                                   & 47.89                                                                               & 19.14                                                                               & 41.52                                                                               & -22.37            & 0.1417              & 1.5054       & 1.0570               & 1.4815                        & 2.27                           \\ 
		\hhline{------------}
		0.970         & 0.8                                       & 6.958                                                                                                                   & 40.65                                                                               & 18.00                                                                               & 38.07                                                                               & -20.07            & 0.1500              & 1.3864       & 0.9831               & 1.9913                        & 2.58                           \\ 
		\hhline{------------}
		0.985         & 0.4                                       & 6.993                                                                                                                   & 39.96                                                                               & 17.11                                                                               & 36.10                                                                               & -18.98            & 0.1475              & 1.3440       & 0.8880               & 1.7721                        & 2.45                           \\ 
	\hhline{------------}
		1.0             & 0                                         & \begin{tabular}[c]{@{}l@{}}7.021, \\6.97 \textbf{\textsuperscript{a}}, \\7.016 \textbf{\textsuperscript{b}}\end{tabular} & \begin{tabular}[c]{@{}l@{}}36.03, \\41.70 \textbf{\textsuperscript{b}} \\ 41.73 \textbf{\textsuperscript{c}}\end{tabular} & \begin{tabular}[c]{@{}l@{}}15.97, \\17.30 \textbf{\textsuperscript{b}} \\ 17.23 \textbf{\textsuperscript{c}}\end{tabular} & \begin{tabular}[c]{@{}l@{}}33.96, \\30.61 \textbf{\textsuperscript{b}} \\ 25.99 \textbf{\textsuperscript{c}} \end{tabular} & -17.99            & 0.1473              & 1.0853       & 0.6672               & 2.0192                        & 2.60                           \\ 
		\hhline{------------}
		1.015         & -0.35                                     & 7.063                                                                                                                   & 31.91                                                                               & 15.26                                                                               & 23.43                                                                               & -8.17             & 0.2026              & 1.0849       & 0.6851               & 1.4036                        & 2.32                           \\ 
	\hhline{------------}
		1.030         & -0.65                                     & 7.098                                                                                                                   & 32.21                                                                               & 14.55                                                                               & 20.79                                                                               & -6.24             & 0.2096              & 1.0883       & 0.6710               & 0.9356                        & 2.01                           \\ 
		\hhline{------------}
		1.045         & -1.3                                      & 7.133                                                                                                                   & 29.83                                                                               & 14.47                                                                               & 18.53                                                                               & -4.06             & 0.2284              & 1.0579       & 0.6919               & 0.9937                        & 2.07                           \\ 
		\hhline{------------}
		1.060         & -1.5                                      & 7.168                                                                                                                   & 32.05                                                                               & 15.44                                                                               & 16.87                                                                               & -1.43             & 0.2482              & 1.0683       & 0.7104               & 0.6287                        & 1.81                           \\
		\hhline{============}
	\end{tabular}

$^{a} $Ref. \cite{ref14}, $ ^{b} $Ref. \cite{ref18}, $ ^{c} $Ref. \cite{ref48}. 
\end{table*}

\subsection{III.B Mechanical properties}

The elastic constants (C$ _{ij} $) are key parameters for the mechanical behaviour of materials, including mechanical stability, anisotropic response, stiffness, and hardness. The computed C$ _{ij} $ and their related mechanical properties, such as Poisson's ratio ($\nu$), Pugh ratio (B/G), machinability index ($\mu_{M}$), and the universal anisotropic index (A$ ^{U} $) under hydrostatic pressures, are reported in Table~\ref{Tab_1}. It is worth mentioning that the studied hydrostatic pressures are of two types, tension and compression, which are addressed by the volume ratios (V/V$ _{0} $) in this table. However, the range of pressures related to volume ratios is also included. The volume ratios in this study have been chosen in a way that allows us to observe changes in mechanical behaviour \textit{(e.g.}, negative Poisson's ratio) as well as topological phase transition (TPT).

Among the three constants C$ _{11} $, C$ _{12} $, and C$ _{44} $, the C$ _{44} $ constant represents resistance to shear deformation, while the C$ _{11} $ relates to unidirectional strain along the principal crystallographic directions. The value of C$ _{44} $ in Table~\ref{Tab_1} is lower than C$ _{11} $ in all pressures, which means that the cubic cell is more easily deformed by a shear in comparison to the unidirectional stress. A material's Cauchy pressure (C$ _{P} $) can also be a useful mechanical parameter, and it is defined as C$ _{P} $ = (C$ _{12} $ - C$ _{44} $) \cite{ref29}. Angular characteristics of atomic bonding as well as the brittleness and ductility of a material can be described by Cauchy pressure \cite{ref29,ref33}. The Cauchy pressure of a ductile material is positive, while the Cauchy pressure of a brittle material is negative. On the other hand, ionic and covalent bonding in a material is correlated with positive and negative Cauchy pressures, respectively. According to Pettifor's rule \cite{ref34}, a material with a large positive C$ _{P} $ will possess significant metallic bonds and exhibit high levels of ductility, while a material that has a negative C$ _{P} $ will possess more angular bonds, resulting in greater brittleness and covalent bonding. Hence, the negative value of C$ _{P} $ predicts that the NaLi$ _{2} $Bi compound is brittle in nature, and some covalent bonding is present in this compound. In addition, these results demonstrate that in tension (compression) hydrostatic pressures, this material approaches a high level of ductility (brittleness).
 \begin{figure*}
	\includegraphics[scale=1.0]{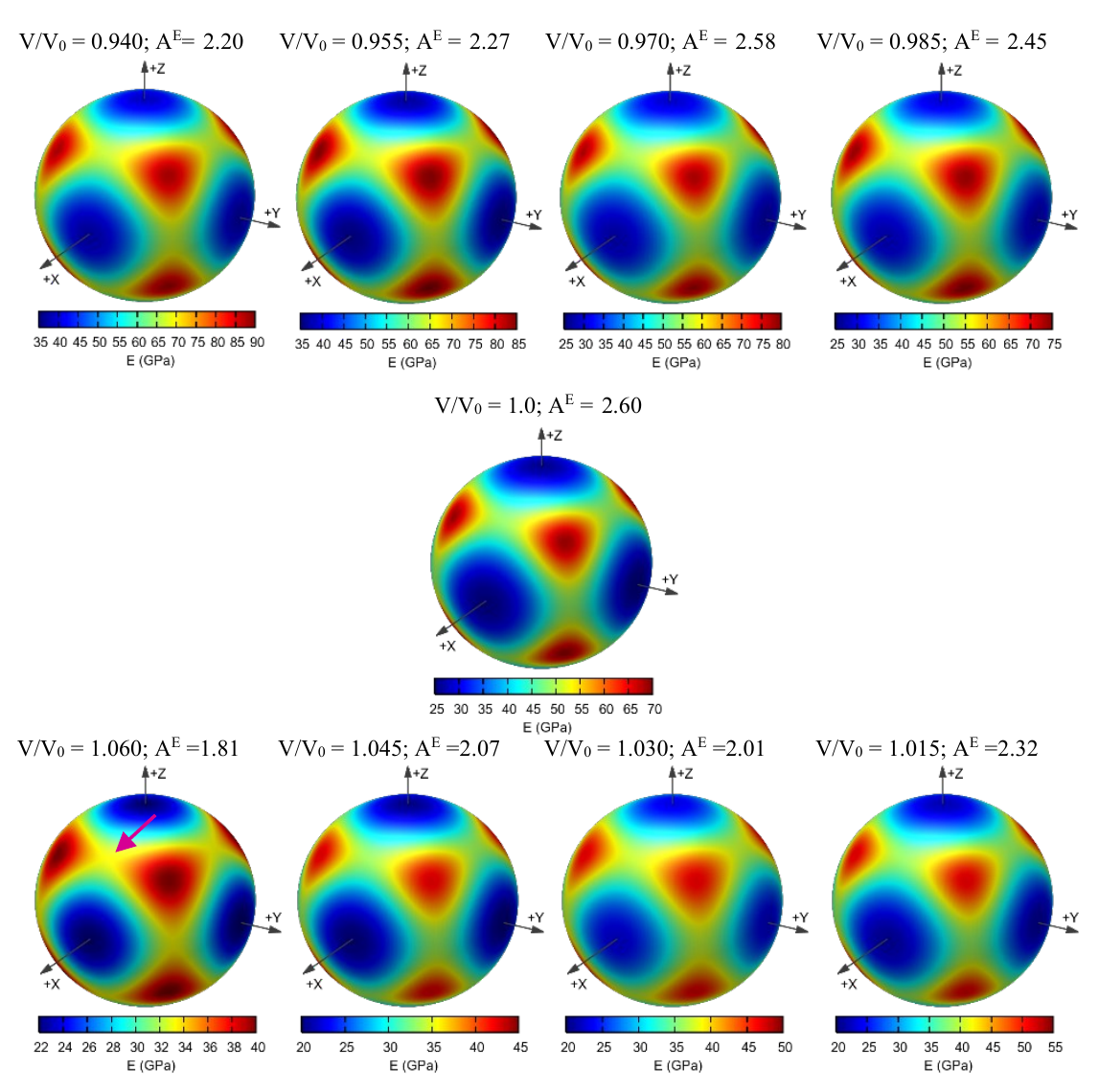} 
	
	\caption{\label{fig:wide_2}3D representation surfaces of Young’s modulus (E) of NaLi$ _{2} $Bi under hydrostatic tension and compression. The more localized the colours and the more complex the colour distribution pattern, the higher the level of anisotropy. The purple arrow in V/V$ _{0} $ = 1.060 shows that the localization in the color’s distribution is reduced.}
	
\end{figure*}
Using the Voigt-Reuss-Hill (VRH) approximation, the Poisson's ratio ($\nu$) and Pugh ratio (B/G) are estimated. Details of this approximation are addressed in Ref. \cite{ref29}. Poisson's ratio and Pugh ratio are two factors that determine a system's brittle and ductile nature in addition to Cauchy pressure. The Poisson's ratio measures deformation (\textit{i.e.}, expansion or contraction) along the perpendicular to the loading direction. Frantsevich \textit{et al.} \cite{ref50} have separated the brittleness and ductility of a material based on Poisson's ratio ($\nu$), defining the border between the two as $\nu$ = 0.26. So, materials with Poisson's ratio values greater and smaller than 0.26 are considered ductile and brittle, respectively. Moreover, in 1954, Pugh has proposed that the bulk modulus to shear modulus ratio (or vice versa) represents a good criterion to identify the brittleness and ductility of materials \cite{ref51}. The critical value that separates brittle and ductile materials is about 1.75. Therefore, materials with a Pugh's ratio greater than 1.75 are predicted to be ductile; otherwise, they are expected to be brittle. Based on these two criteria, it can be seen that the NaLi$ _{2} $Bi compound has brittle in nature in the equilibrium state (V/V$ _{0} $ = 1.0), and this characteristic is preserved in the studied hydrostatic pressures. It is noteworthy that a significant agreement exists between the two conditions related to $\nu$ and B/G, and the results associated with C$ _{P} $. Poisson's ratio is more sensitive to the detection of physical characteristics than two C$ _{P} $ and B/G parameters. The Poisson's ratio also provides information about the bonding nature of the materials. Materials that are covalently bonded tend to have Poisson's ratios around 0.1, whereas materials that are ionic tend to have Poisson's ratios around 0.25 \cite{ref52}. Based on Poisson's ratio values, it is predicted that most of the chemical bonds in this composition are covalent, and the covalent bonds become more intense at compression hydrostatic pressures. Also, in tensile hydrostatic pressures, ionic chemical bonds play a prominent role compared to covalent bonds. A Poisson's ratio can provide insight into the nature of interatomic forces in solids \cite{ref53, ref54}. Materials that have Poisson's ratios within 0.25 and 0.50 are called central force solids, whereas non-central force solids have Poisson's ratios either less than 0.25 or greater than 0.50 \cite{ref54}. In the NaLi$_{2}$Bi compound, therefore, non-central forces of atoms prevail, while in tension hydrostatic pressures, central forces become more important (see Table~\ref{Tab_1}).

Another useful mechanical performance index is the machinability index ($\mu_{M}$), and it is defined as $\mu_{M}$ = B/C$ _{44} $ (B is bulk modulus). This parameter measures a material's machinability, or its ability to be machined using cutting or shaping tools. A material's degree of machinability is becoming increasingly important to the industry today because it determines factors such as how much power is needed for cutting, what temperature to use, and how much plastic strain to apply. Furthermore, it can be used to determine a solid's plasticity and dry lubricating ability \cite{ref35,ref36}. Having a high $\mu_{M}$ value indicates excellent lubricating properties, which results in lower friction. The value of $\mu_{M}$ (= 0.6672) indicates that the NaLi$ _{2} $Bi compound has a good level of machinability. Under hydrostatic compression (V/V$ _{0} $ < 1), however, this characteristic increases, indicating it could be an attractive feature for machine tool applications.
 \begin{figure*}
	\includegraphics[scale=1.0]{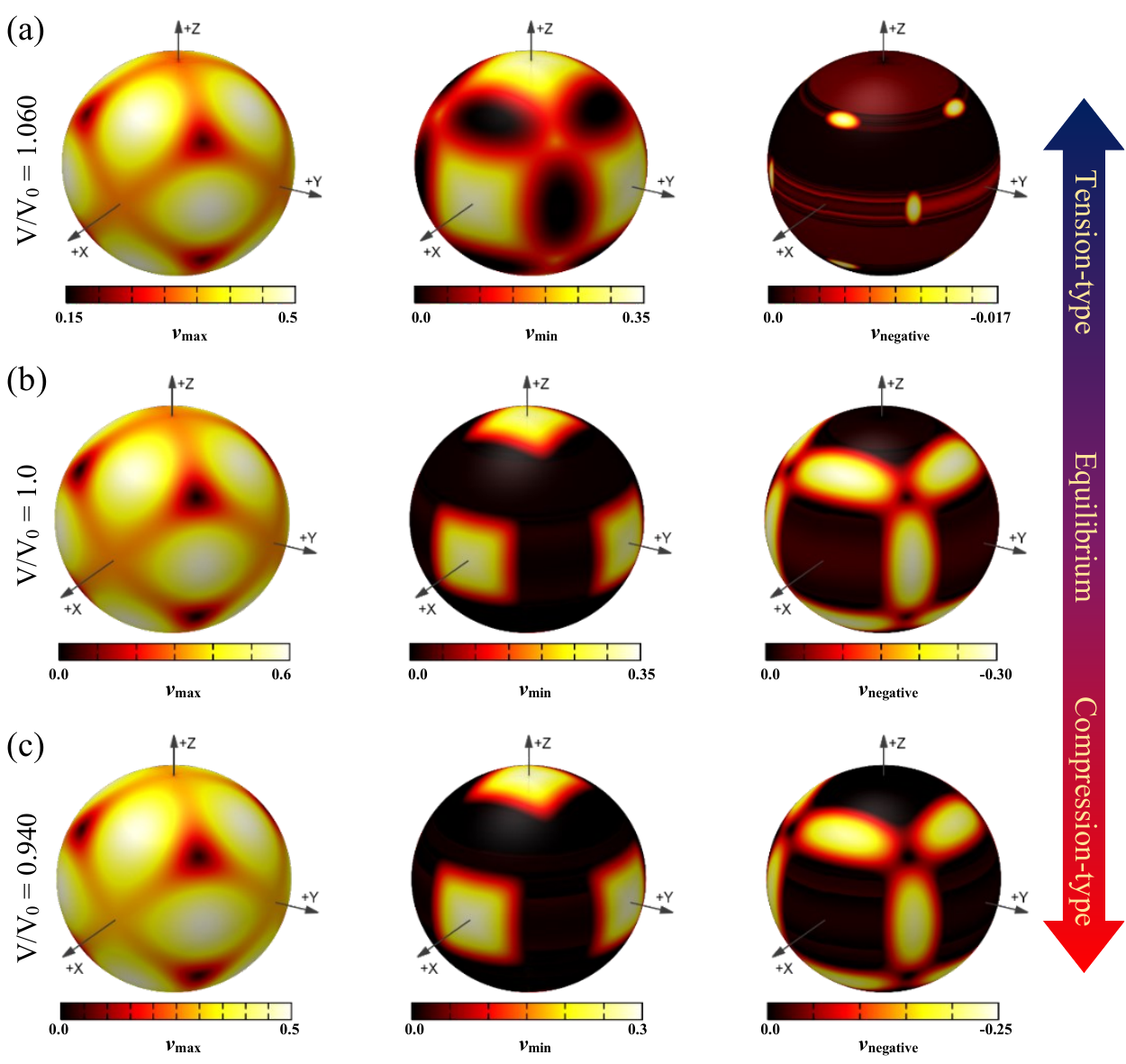} 
	\caption{\label{fig:wide_3}3D representation surfaces of Poisson's ratio for NaLi$_{2}$Bi in (a) V/V$_{0}$ = 0.940 , (b) V/V$_{0}$ = 1.0, and (c) V/V$_{0}$ = 1.060 states. The maximum positive, minimum positive, and negative values of Poisson's ratio are $\nu_{max}$, $\nu_{min}$, and $\nu_{negative}$, respectively. The maximum and minimum values of Poisson's ratio are located at [110]([101] or [011]) and [100]([001] or [010]) directions, respectively.}
	
\end{figure*}

Plastic deformation in crystals, the formation of microscale cracks in ceramics, and plastic relaxation in thin films are all influenced by elastic anisotropy. Hence, evaluating elastic anisotropy factors is essential for predicting solid behaviour under different external stress conditions. The universal anisotropy index (A$ ^{U} $) and the anisotropy of Young's modulus (A$ ^{E} $) of NaLi$ _{2} $Bi are listed in Table~\ref{Tab_1} (the calculation of this two indices is detailed in Ref. \cite{ref29}). In V/V$ _{0} $ = 1.0, there is the highest degree of anisotropy (A$ ^{U} $ = 2.0192). According to the trend of changes in this index under hydrostatic tension and compression, it can be found that the degree of anisotropy (\textit{i.e.}, A$ ^{U} $) decreases slowly in tension-type of pressures and more rapidly in compression-type of pressures. Young's modulus (E) can also be a good indicator of the level of anisotropy. The 3D representation surfaces of Young's modulus are calculated by \textsc{ElaTools} and included in Fig.~\ref{fig:wide_2}. The more localizing colors and the more complex the pattern of color distribution in this figure, the higher the level of anisotropy is. According to A$ ^{E} $ in Table~\ref{Tab_1} and Fig.~\ref{fig:wide_2}, it can be seen that NaLi$ _{2} $Bi has high anisotropy. With increasing tension-type pressure, anisotropy decreases, and its minimum value corresponds to a state with V/V$ _{0} $ = 1.060 and A$ ^{E} $ = 1.81. This is due to the reduced localization in the colors distribution pattern (see purple arrow).

 \begin{figure*}
	\includegraphics[scale=0.9]{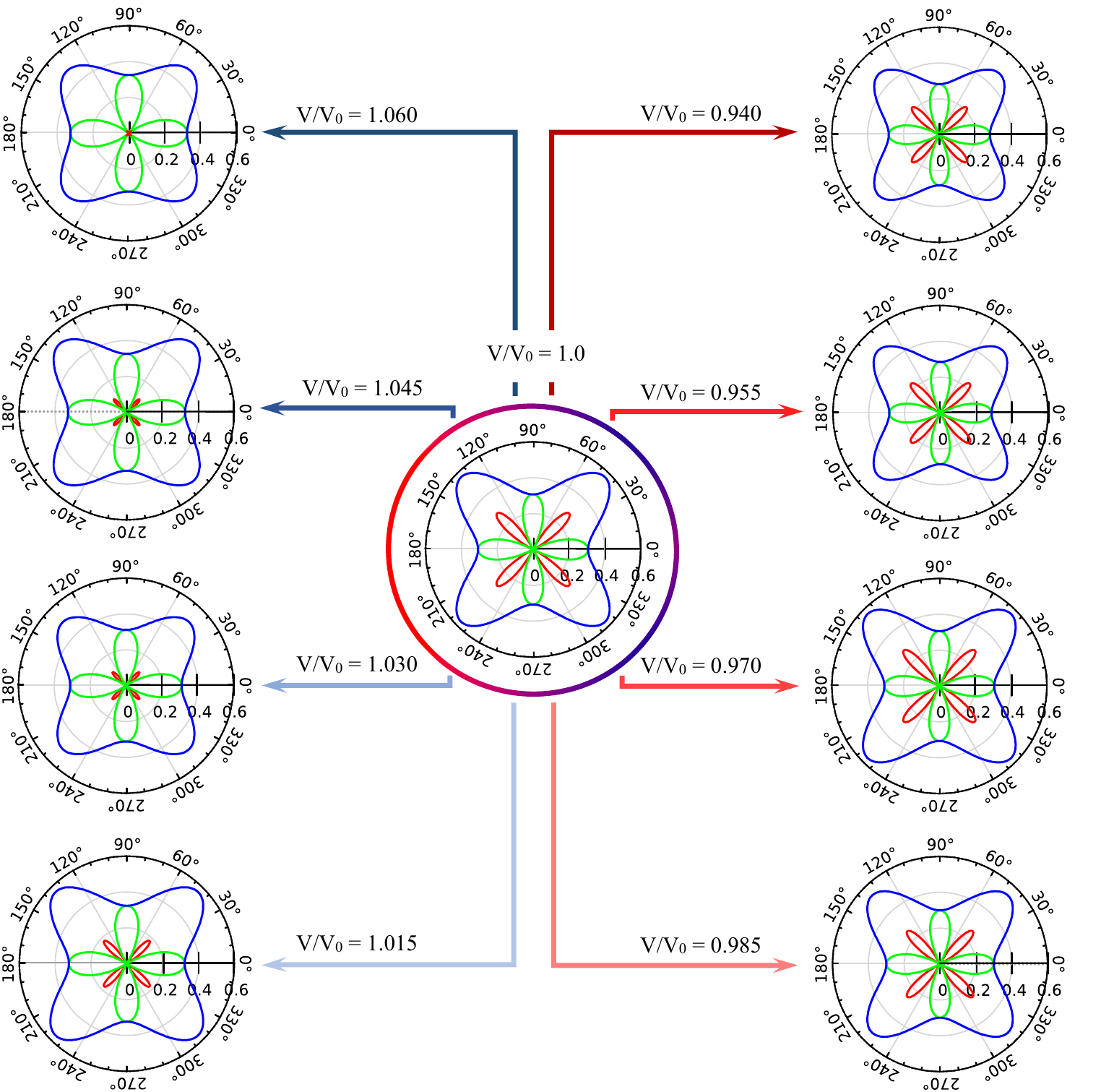} 
	
	\caption{\label{fig:wide_4}2D polar representation of Poisson's ratio in the (100)-plane for NaLi$ _{2} $Bi in equilibrium state (V/V$ _{0} $ =1.0 ), and hydrostatic tension (V/V$ _{0} $ > 1 ), and compression (V/V$ _{0} $ < 1). Blue (green) and red colors represent the maximum (minimum) positive and negative values of Poisson's raito, respectively. With the increase of hydrostatic tension (V/V$ _{0} $ > 1), the value of NPR decreases until the critical point V/V$ _{0} $ = 1.060, while in hydrostatic compression (V/V$ _{0} $ < 1), the value of NPR remains almost constant.}
	
\end{figure*}
Materials belonging to the \textit{auxetic} family possess negative Poisson's ratio (NPR), causing expansion under tension rather than contraction. So, this enables the development of innovative technologies for impact-resistant composites, ultra-precise sensors, tougher ceramics, and high-performance armor due to auxetic materials' atypical elastic behaviour \cite{ref37,ref38}. This feature is investigated using 3D surface representations (see Fig.~\ref{fig:wide_3}) and 2D polar plots (see Fig. ~\ref{fig:wide_4}) of Poisson's ratio of NaLi$ _{2} $Bi in (100)-plane under hydrostatic tension and compression. Based on Fig.~\ref{fig:wide_3}, the NaLi$ _{2} $Bi compound has a NPR of $\nu_{negative}$ = -0.285 along the [110]([101] or [011])-direction in the equilibrium state. As a result, this compound can be regarded as an \textit{auxetic} material. The results show that with increasing hydrostatic tension, the negative Poisson's rate decreases and reaches a critical value of -0.017 (see Fig.~\ref{fig:wide_3}(a)). From this point on, the compound loses its \textit{auxetic} behaviour (see Fig. \textcolor{red}{2S}). Therefore, a type of mechanical phase transition (MPT) from the \textit{auxetic} state to the \textit{non-auxetic} state has occurred for this compound. Fig.~\ref{fig:wide_4} displays the 2D polar representation of Poisson’s ratio in the (100)-plane for the studied pressures. The blue (green) represents the maximum (minimum) value and the red represents the negative value. In this figure, it can be clearly seen that with the increase of hydrostatic tension (V/V$ _{0} $ > 1), the value of NPR decreases to the critical point of V/V$ _{0} $ = 1.060. In hydrostatic compression (V/V$ _{0} $ < 1), the value of NPR remains almost constant. It is noteworthy that the maximum and minimum values of Poisson's ratio are located at [110]( [101] or [011]) and [100]([001] or [010]) directions, respectively.

\subsection{III.C Topological properties}
 \begin{figure*}
	\includegraphics[scale=1.0]{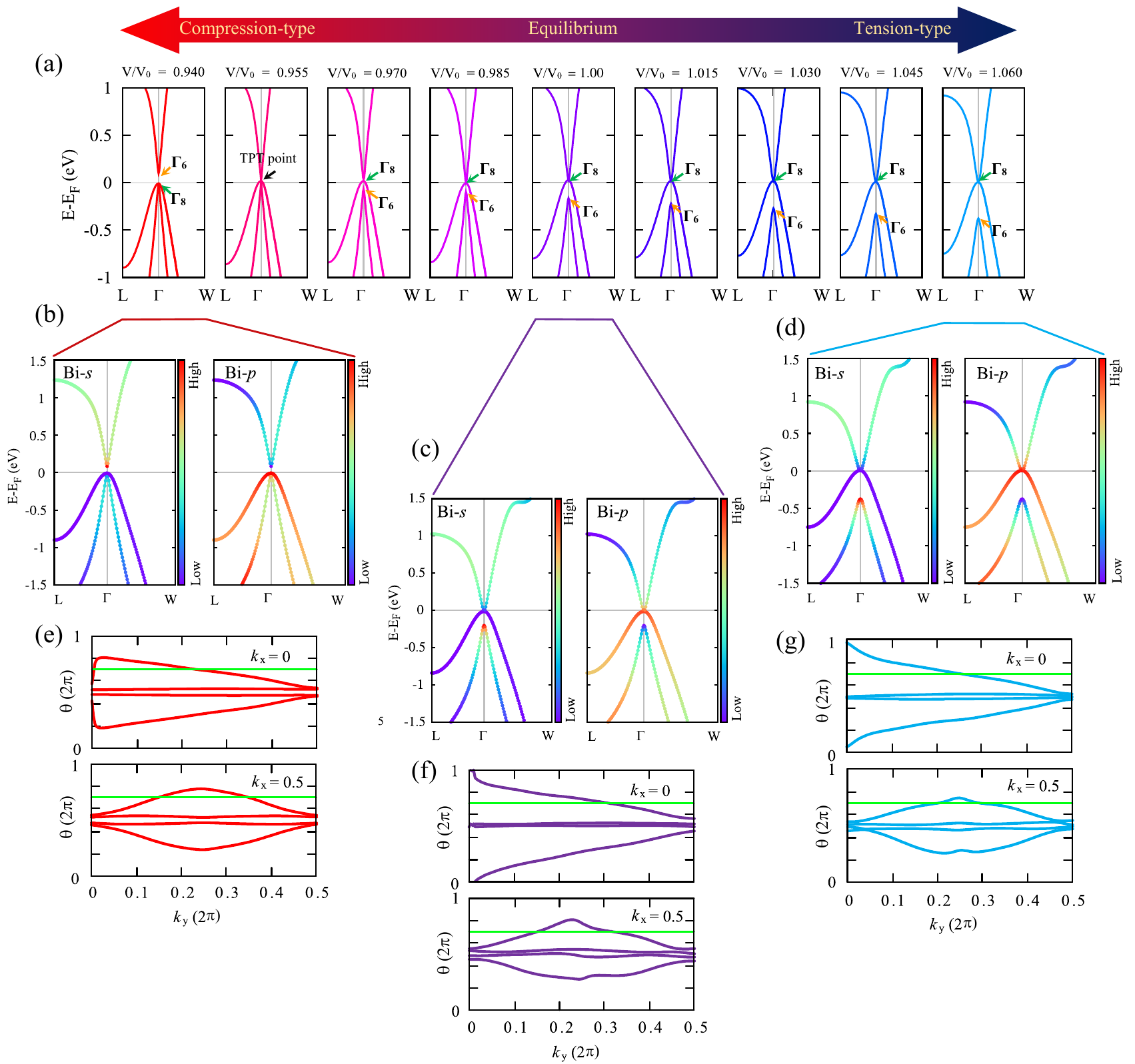} 
	
	\caption{\label{fig:wide_5} (a) The band structures of the NaLi2Bi compound under hydrostatic tension (V/V$_{0}$ > 1 ), compression (V/V$_{0}$ < 1)  , and equilibrium states (V/V$_{0}$ = 1.0) using the PBE-GGA approach. The $\Gamma_{6}$ (Bi-\textit{s}-liked) and $\Gamma_{8}$ (Bi-\textit{p}-liked) states are marked with orange and green arrows, respectively. The black arrow shows the critical point of the topological phase transition in the V/V$_{0}$ = 0.955 state. The orbital projected band structures of the NaLi$_{2}$Bi in the (b) V/V$_{0}$ = 0.940, (c) equilibrium state (V/V$_{0}$ = 1.0), and (d) V/V$_{0}$ = 1.060 along high symmetry line L- $\Gamma$-W. The evolution lines of the Wannier charge centers along $k_{y}$ in the $k_{x}$ = 0 and $k_{x}$ = 0.5 planes in the (e) V/V$_{0}$ = 0.940, (f) V/V$_{0}$ = 1.0, and (g) V/V$_{0}$ = 1.060 states. The arbitrary reference line (green line) is shown for $k_{x}$ = 0 and $ k_{x} $ = 0.5 planes.}
	
\end{figure*}

\begin{figure*}
	\includegraphics[scale=1.0]{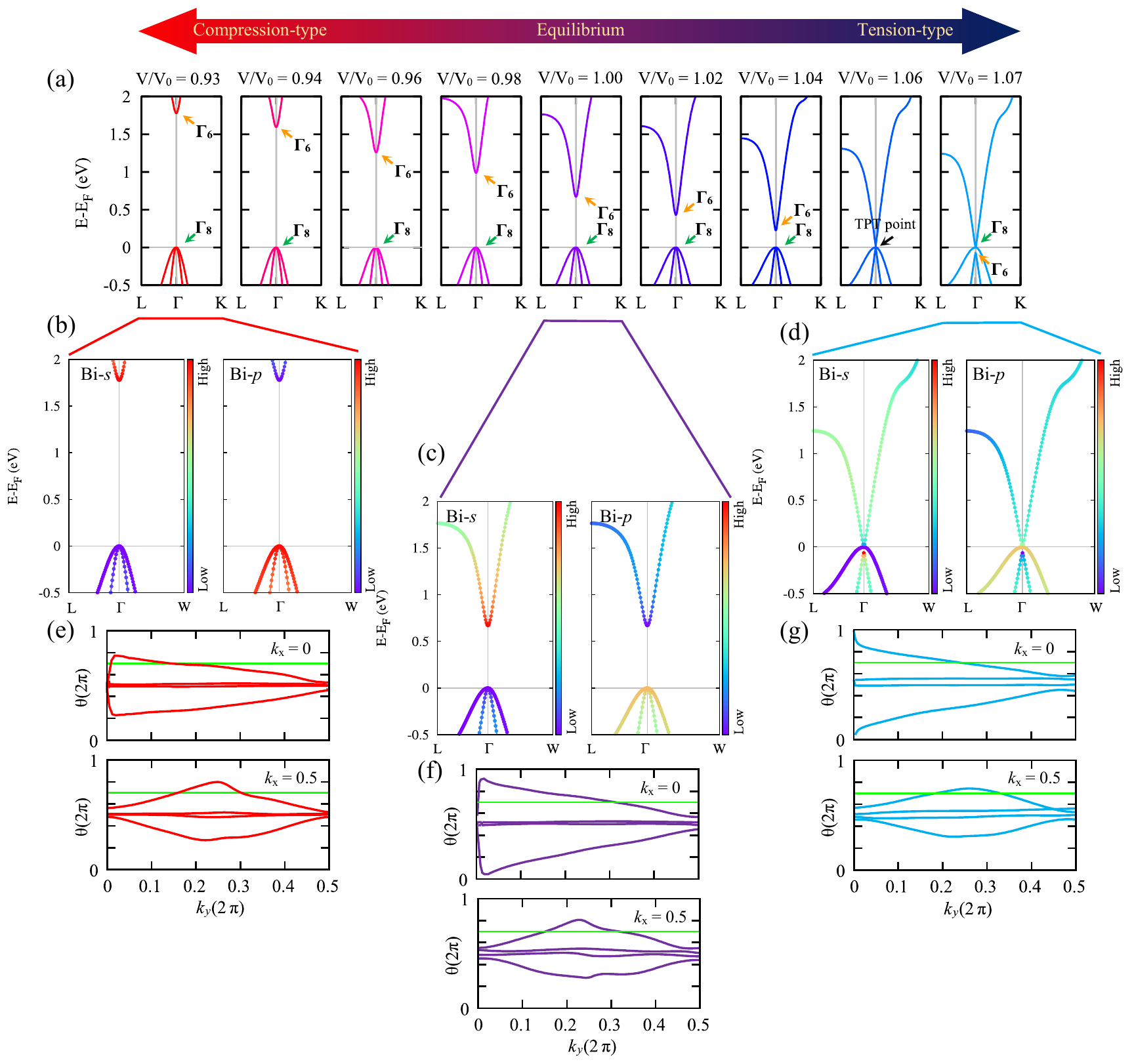} 
	
	\caption{\label{fig:wide_6} (a) The band structures of the NaLi2Bi compound under hydrostatic tension (V/V$_{0}$ > 1 ), compression (V/V$_{0}$ < 1 )  , and equilibrium states (V/V$_{0}$ = 1.0) using the TB-mBJ approach. The $\Gamma_{6}$ (Bi-\textit{s}-liked ) and $\Gamma_{8}$ (Bi-\textit{p}-liked) states are marked with orange and green arrows, respectively. The black arrow shows the critical point of the topological phase transition in the V/V$ _{0} $ = 1.06 state. The orbital projected band structures of the NaLi$ _{2} $Bi in the (b) V/V$ _{0} $ = 0.930, (c) equilibrium state (V/V$ _{0} $ = 1.0), and (d) V/V$ _{0} $ = 1.070 along high symmetry line L- $\Gamma$-W. The evolution lines of the Wannier charge centers along $ k_{y} $ in the $ k_{x} $ = 0 and $ k_{x} $= 0.5 planes in the (e) V/V$ _{0} $ = 0.930, (f) V/V$ _{0} $ = 1.0, and (g) V/V$ _{0} $= 1.070 states. The arbitrary reference line (green line) is shown for $ k_{x} $= 0 and $ k_{x} $ = 0.5 planes.}
	
\end{figure*} 

Topological insulators (TIs) with novel surface states have recently drawn considerable theoretical and experimental attention \cite{ref21,ref39,ref40}. The spin-orbit coupling (SOC) and band inversion at the Fermi energy give TI remarkable electronic properties. This makes TI ideal for applications in electronics, thermoelectrics, optoelectronics, superconductors, and magnetics \cite{ref40,ref41,ref42,ref43}. It has been found that it is possible to control topological band order using electromagnetic fields, alloying, pressure, strain, and temperature effects \cite{ref7,ref20,ref44,ref45, ref59}. Here, the mechanism of hydrostatic strain is employed in this study in order to investigate the TPT of the NaLi$ _{2} $Bi compound. Here, the mechanism of hydrostatic strain is employed in this study in order to investigate the TPT of the NaLi$_{2}$Bi compound using PBE-GGA and TB-mBJ approaches.

•	\textit{PPE-GGA approach:} To investigate topological band order, we first examine the equilibrium state of the NaLi$ _{2} $Bi compound. Fig. \textcolor{red}{3S} illustrates the orbital projected band structure of NaLi$ _{2} $Bi without SOC at an equilibrium state. Compared to Ref. \cite{ref14}, the bandgap of this compound along high symmetry line L-$\Gamma$-W is 0.41 eV, which is in good agreement with this reference (0.4 eV). In this figure, based on the orbital projected bands (fat bands), it is given that the Bi-\textit{s}-liked ($\Gamma_{1}$) is above Bi-\textit{p}-liked ($\Gamma_{5}$) at $\Gamma$ point. It can be concluded that in the absence of SOC, the band inversion did not occur and thus the compound is a semiconductor with an trivial bandgap. This configuration is similar to HgTe compound in the absence of SOC \cite{ref55}. When the SOC is included, the bandgap is closed and Bi-\textit{s}-liked orbitals are exchanged with Bi-\textit{p}-liked (Bi-\textit{s}-liked below Bi-\textit{p}-liked band), and a band inversion of \textit{sp}-type occurs (see Fig.~\ref{fig:wide_5}(a and c)). In this case  (V/V$_{0}$ = 1.0), the compound is reduced to a zero-gap semimetal with an inverted band order. This is the result of familiarity, which is created due to the presence of strong SOC in TIs and topological semimetals  \cite{ref8,ref55, ref56}. In cubic crystals and in the presence of  SOC, the $\Gamma$ point around Fermi level can be traced back to the \textit{s}-like doublet $\Gamma$$ _{6} $ (\textit{j} = 1/2) state, the \textit{p}-like quartet $\Gamma$$ _{8} $ (\textit{j} = 3/2) state, and the split-off doublet $\Gamma$$ _{7} $ state (\textit{j} = 1/2) (($\Gamma _{7}$ state is indicated in Fig. \textcolor{red}{4S})) \cite{ref46}. Here "\textit{j}" is the total angular momentum. In fact, SOC splits the $\Gamma_{5}$ state (see Fig. \textcolor{red}{3S} in the absence of SOC) into $\Gamma_{8}$ (\textit{j} = 3/2, denoted as \textit{p}$_{3/2}$) and $\Gamma_{7}$ (\textit{j}= 1/2, denoted as \textit{p}$_{1/2}$) states \cite{ref55,ref56}. Based on the energy difference between the $\Gamma _{8} $ and $\Gamma _{6} $ states in the band structure, a parameter can be defined (\textit{i.e.}, $\delta$ = [$\Gamma _{6} $ -$\Gamma$$ _{8} $]) that shows the band inversion behaviour, which is known as the band inversion strength (BIS). Therefore, for nontrivial topological phases, this value is negative ($\delta$ < 0) and for trivial topological phases, this value is positive ($\delta$ > 0). Also, this value becomes zero for the TPT point. In order to investigate TPT, the evolution of the band structure at the $\Gamma$ point under hydrostatic tension (V/V$ _{0} $ > 1) ,and compression (V/V$ _{0} $ < 1) has been evaluated in Fig. \ref{fig:wide_5}(a). In addition, in this figure, $\Gamma_{6}$ and $\Gamma_{8}$ states are marked with orange and green arrows, respectively. It can be found that when hydrostatic pressures of the tension type are applied, the nontrivial topological phase is preserved and the |$\delta$| ($\delta$ < 0) also increases. For example, the orbital projected band structure in the V/V$ _{0} $ = 1.060 state from Fig.~\ref{fig:wide_5}(d) shows that there is the \textit{sp}-band inversion and the only difference with the equilibrium state is in the value of the $\delta$. Meanwhile, with the increase of the hydrostatic pressure of the compression type, the $\delta$ decreases and reaches zero value in V/V$ _{0} $ = 0.955 state. This state shows the critical point of the topological phase transition, which is indicated by the black arrow. The value of critical pressure, in this case, is estimated as 1.0 GPa, according to Table~\ref{Tab_1}. It is worth noting that the TPT is from the nontrivial to the trivial state. When we cross the critical point of the TPT, at V/V$ _{0} $ = 0.940, the Bi-\textit{p}-liked ($\Gamma_{8}$) is below the Bi-\textit{s}-liked ($\Gamma_{6}$) and the compound is converted into a semiconductor with a trivial bandgap of $\approx$ 0.1 eV (see Figs.~\ref{fig:wide_5}(a and b)).

\begin{figure*}
	\includegraphics[scale=0.88]{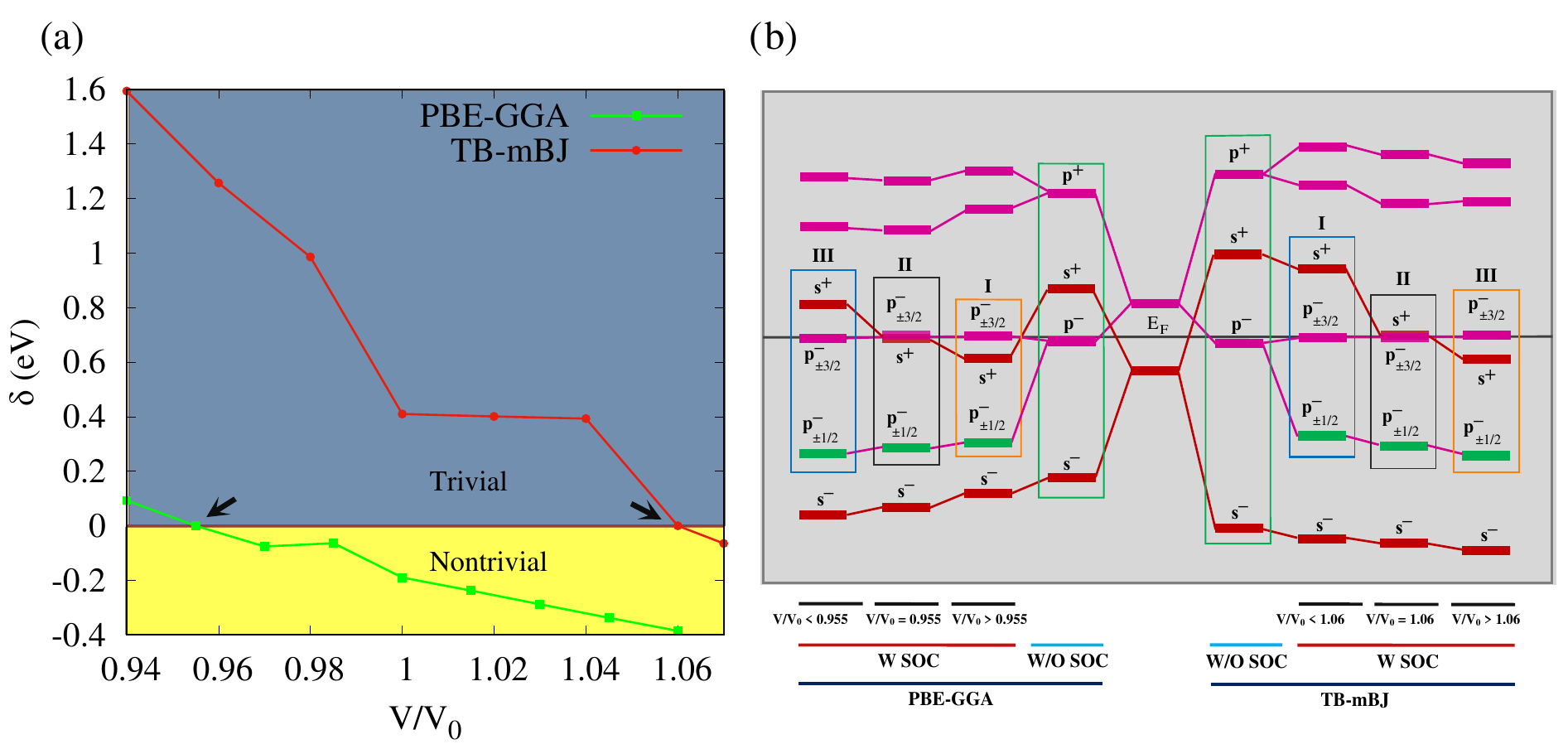} 
	\caption{\label{fig:wide_7} (a) Band inversion strength ($\delta$) of NiLi$_{2}$Bi compound in two approaches, PBE-GGA and TB-mBJ, as a function of V/V$_{0}$. The black arrows show the topological phase transitions (TPT) points. (b) Schematic evolution of the atomic energy levels at $\Gamma$ point under hydrostatic pressure with (denoted as W SOC) and without SOC (denoted as W/O SOC) within PBE-GGA and TB-mBJ approaches. The symbols "+" and "-" label the parities of the bands at the $\Gamma$ point. Orange, black, and green rectangles represent nontrivial states, TPT points, and trivial states, respectively.}
	
\end{figure*}

To confirm that the topological phase transition occurs in this structure under hydrostatic tension and compression, we calculated the topological $\Bbb Z_{2}$-index by the evolution lines of Wannier Charge Centres (WCC) along $ k_{y} $ in the two time-reversal invariant planes $ k_{x} $= 0 and $\pi$ in the Brillouin zone, as shown in Figs.~\ref{fig:wide_5}(e-g). For  V/V$ _{0} $ > 0.955 states, an odd number of points in the $ k_{x} $=0 plane are crossed by the evolution curve in any arbitrary line horizontal axis (green line), while an even number of points in the $ k_{x} $= 0.5 plane are crossed (\textit{e.g.}, Figs.~\ref{fig:wide_5}(f and g)); thus yielding (z$ _{2} $)$ _{0} $ = 1, (z$ _{2} $)$ _{0.5} $ = 0 and $\Bbb Z_{2}$  = [((z$ _{2} $)$ _{0} $ + (z$ _{2} $)$ _{0.5} $) mod 2] = 1. Accordingly, NaLi$ _{2} $Bi is a nontrivial topological zero-gap semimetal with a topological index $\Bbb Z_{2}$  = 1 for V/V$ _{0} $ > 0.955. In the same way, for the states that are after the TPT point V/V$ _{0} $ < 0.955  (\textit{e.g.}, V/V$_{0}$ = 0.940 state in Fig. 5(e)), the topological index becomes $\Bbb Z_{2}$  = 0 ((z$ _{2} $)$ _{0} $ = 0, (z$ _{2} $)$ _{0.5} $ = 0 and $\Bbb Z_{2}$  = [((z$ _{2} $)$ _{0} $ + (z$ _{2} $)$ _{0.5} $) mod 2] = 0), which shows that the compound has transitioned to the trivial topological phase.

•	\textit{TB-mBJ approach:} It is well known that TB-mBJ gives a very well estimate of the band gap of solids as is the case for the GW method and the hybrid functional, but with a less cost \cite{ref57, ref49}. In Fig.~\ref{fig:wide_6}, we re-performed the calculations related to Fig.~\ref{fig:wide_5} (PBE-GGA approach) in the TB-mBJ approach to better understand the topological properties of the studied compound. In the equilibrium state (V/V$_{0}$ = 1.0), the compound has a trivial bandgap (0.62 eV) in the presence of SOC, as shown in Fig.~\ref{fig:wide_6}(a). As can be seen in Fig.~\ref{fig:wide_6}(c), this compound is a conventional (trivial topological semiconductor) semiconductor since the band inversion has not occurred (Bi-\textit{p}-liked below Bi-\textit{s}-liked band). The calculations of the topological $\Bbb Z_{2}$-index ($\Bbb Z_{2}$ = 0) performed by the evolution lines of WCC are also a confirmation of this issue (see Fig.~\ref{fig:wide_6}(f)). With the TB-mBJ approach, in contrast to the PBE-GGA approach, where the TPT occurs under hydrostatic compression, the trivial bandgap increases during hydrostatic compression and the compound maintains its trivial topological phase as a result (see Fig.~\ref{fig:wide_6}(a)). For example,  Figs.~\ref{fig:wide_6}(b and e) show the orbital projected band structures and WCC at V/V$_{0}$ = 0.93, which indicates that the band inversion did not occur and that the topological $\Bbb Z_{2}$-index is zero ($\Bbb Z_{2}$ = 0). As can be seen in Fig.~\ref{fig:wide_6}(a), with the increase of the hydrostatic pressure of the tension-type, the bandgap decreases and in the V/V$_{0}$ = 1.06 state, we reach the critical point of the TPT (see black arrow) where the BIS is zero. The hydrostatic pressure of this critical point is -3.4 GPa (in this literature tensile hydrostatic pressure is indicated by the negative sign). It is at this critical point that the TPT from a trivial to a nontrivial state occurs. A band inversion of \textit{sp}-type has occurred above the critical pressure (\textit{i.e.}, V/V$_{0}$ = 1.07 state), as exhibited in Fig.~\ref{fig:wide_6}(a and d), and the BIS has become negative. The compound in this case results in a nontrivial topological zero-gap semimetal. As seen in Fig.~\ref{fig:wide_6}(g), the evolution lines of WCC also show that the topological $\Bbb Z_{2}$-index is nonzero ( $\Bbb Z_{2}$ = 1).

In Fig.~\ref{fig:wide_7}(a), we can compare the TPT of PBE-GGA and TB-mBJ approaches based on how the band inversion strength ($\delta$) changes as a function of V/V$_{0}$. The yellow and blue areas represent $\delta$ < 0  (nontrivial state) and $\delta$ > 0 (trivial state), respectively. As seen, in PBE-GGA approach, TPT point corresponds to hydrostatic compression (V/V$_{0}$ < 1.0), while in TB-mBJ approach, it corresponds to hydrostatic tension (V/V$_{0}$ > 1.0) (see black arrows). It should be noted that the TPT point in the PBE-GGA approach is from a nontrivial state to a trivial state, while it is reversed in the TB-mBJ approach. It is possible to define $\delta_{SOC}$ (= [$\Gamma_{8}$ - $\Gamma_{7}$]) as another parameter similar to $\delta$, which indicates the strength of SOC. The changes of this parameter within PBE-GGA and TB-mBJ approaches are shown in Fig. \textcolor{red}{4S}. The results obtained in two approaches show that this parameter increases (decreases) with the increase of hydrostatic pressure of the compression (tensile) type.

To explain the physical mechanism underlying the band inversion and the parity exchange of this compound, the schematically evolution of the energy levels of the atomic levels at $\Gamma$ point under hydrostatic pressure (compression and tension types), in the presence and absence of SOC for the two approaches studied, is shown in Fig. ~\ref{fig:wide_7}(b). As per the normal band flling order near Fermi energy, the valence band consists of Bi-\textit{p} orbitals, while the conduction band is composed of \textit{s}-type orbitals (see green rectangle), which, due to the formation of chemical bonds in this compound, hybridize and split into bonding and antibonding states in the absence of SOC (\textit{i.e.}, \textit{p}$^{+}$ (\textit{s}$^{+}$) and \textit{p}$^{-}$ (\textit{s}$^{-}$) with different parities) \cite{ref7, ref58}. The system is therefore an conventional semiconductor with trivial bandgap as shown in Fig.~\ref{fig:wide_7}(b) in both PBE-GGA and TB-mBJ approaches. In the PBE-GGA approach, when the SOC is taken into account under hydrostatic pressures, V/V$_{0}$ > 0.955, (see orange rectangle of stage (I)) the \textit{p}$^{-}$ state splits into the |\textit{p}$^{-}$, \textit{$\pm$ j}〉 with a total angular momentum of \textit{j} = 3/2 and \textit{j} = 1/2. On the other hand, the energy band gap reaches zero and the compound becomes a topological semimetal system. In this case, the parity of \textit{s} and \textit{p}$_{\pm 3/2}$  states near the Fermi energy have been exchanged and band inversion occurs (s$^{+}$ state below \textit{p}$^{-}_{\pm 3/2}$ state). At the critical point (V/V$_{0}$ = 0.955) from stage (II) (see black rectangle), these two states are close together and at the V/V$_{0}$ < 0.955, their parity is exchanged with respect to the stage (I), and the system transitions from the nontrivial state to the trivial state (see blue rectangle of stage (III)). There is a similar mechanism in the TB-mBJ approach, but the only difference is that during stage (I), the compound has a trivial bandgap, and after reaching the TPT point, the parity of \textit{s} and \textit{p}$_{\pm 3/2}$  states is exchanged and the compound becomes a semimetal material with nontrivial topological phase (from the trivial state to the nontrivial state).

\section{I.V. Summary and outlook}
In summary, the mechanical properties and topological phase transition of bi-alkali pnictogen NaLi$ _{2} $Bi are investigated in the PBE-GGA approach from the first-principles calculation. Calculations based on phonon modes and elastic constants indicate that this compound is stable the studied hydrostatic tension and compression. The results show that applying hydrostatic tension leads to a kind of mechanical phase transition in the compound so that the auxetic behaviour of the material changes to \textit{non-auxetic} behaviour after the critical pressure of -1.5 GPa (V/V$ _{0} $ = 1.060). According to the Cauchy pressure, Poisson's ratio, and Pugh ratio, it has been determined that this compound has a brittle behaviour in the equilibrium state, and this behaviour is maintained under the hydrostatic pressures studied in this work. 
Based on the study of the electronic band structure in the equilibrium state, it can be concluded that the NaLi$ _{2} $Bi compound has a zero-gap semimetal in the PBE-GGA approach, in which the sp-band inversion has occurred, and its topological index is $\Bbb Z_{2}$ = 1. On the other hand, these calculations show that, in tension-type hydrostatic pressures, the topological phase transition occurs from the non-trivial phase to the trivial phase, and the critical pressure is $\approx$1 GPa. Additionally, the TB-mBJ approach indicates that this compound is a semiconductor with a trivial bandgap. It is possible to close the bandgap by increasing the tension-type hydrostatic pressure by $\approx$-3.4 GPa, and after passing through this point the compound transforms from a trivial phase into a topological nontrivial phase and becomes a zero-gap semimetal with an inverted band structure (\textit{sp}-band inversion occurs). By comparing the two studied approaches in the equilibrium state of the NaLi$_{2}$Bi compound, it can be concluded that conventional approaches like GGA may provide false results in the detection of topological phases and band inversion in materials. The final point we suggest in this paper is that uniaxial strain may induce other topological classes suitable for practical application in this compound, which deserves further study.

\section{acknowledgement}
~
\section{References}

	\bibliographystyle{apsrev4-1} 
	\bibliography{ref.bib}


\end{document}